\documentclass[lettersize,journal]{IEEEtran}
\usepackage{amsmath,amsfonts}
\usepackage[T1]{fontenc}
\usepackage{algorithm}
\usepackage{algpseudocode}
\usepackage{caption}
\usepackage{gensymb}
\usepackage{tabularray}
\usepackage{array}
\usepackage[caption=false,font=normalsize,labelfont=sf,textfont=sf]{subfig}
\usepackage{textcomp}
\usepackage{stfloats}
\usepackage{url}
\usepackage{verbatim}
\usepackage{graphicx}
\usepackage{cite}
\usepackage{svg}
\usepackage{hyperref}
\usepackage{physics} 
\usepackage{tikz}
\usepackage{quantikz}
\usepackage{xcolor}
\usetikzlibrary{shapes.geometric, arrows.meta, positioning, fit, calc, backgrounds}

\usepackage{listings}
\usepackage{xcolor}
\definecolor{oneLightBackground}{RGB}{250, 250, 250}
\definecolor{oneLightGutter}{RGB}{157, 157, 159}
\definecolor{oneLightText}{RGB}{56, 58, 66}
\definecolor{oneLightKeyword}{RGB}{166, 38, 164}   
\definecolor{oneLightFunction}{RGB}{64, 120, 242}  
\definecolor{oneLightString}{RGB}{80, 161, 79}     
\definecolor{oneLightNumber}{RGB}{152, 104, 1}     
\definecolor{oneLightComment}{RGB}{160, 161, 167}  
\definecolor{oneLightCyan}{RGB}{1, 132, 188}      
\definecolor{oneLightAttribute}{RGB}{228, 86, 73} 
\definecolor{oneLightClass}{RGB}{193, 132, 1}     

\definecolor{netsquidblue}{RGB}{210, 225, 245}
\definecolor{netsquidgreen}{RGB}{220, 240, 220}
\definecolor{netsquidpurple}{RGB}{235, 220, 245}
\definecolor{netsquidorange}{RGB}{255, 235, 200}
\definecolor{darkblue}{RGB}{40, 60, 120}

\definecolor{spdcblue}{RGB}{220, 230, 255}
\definecolor{entorange}{RGB}{255, 240, 210}
\definecolor{hergreen}{RGB}{220, 255, 220}
\definecolor{postred}{RGB}{255, 220, 220}

\newcommand{\qfunc}[1]{{\small\sffamily\bfseries\color{blue!80!black}#1}}
\newcommand{\func}[1]{{\small\sffamily\itshape\color{black!80}#1}}
\newcommand{\cfunc}[1]{{\small\sffamily\bfseries\color{red!80}#1}}

\tikzset{
    layer/.style={
        rectangle, draw=black!70, thick,
        text width=4.9cm, minimum height=1.4cm,
        align=left, font=\sffamily, fill=white,
        inner sep=8pt
    },
    control/.style={
        rectangle, draw=orange!80, thick, fill=orange!10,
        text width=2.2cm, align=center, font=\sffamily
    },
    operator/.append style={fill=red!60},
    phase/.append style={fill=blue!60},
    xgate/.style={fill=blue!10, draw=blue!50!black},
    egate/.style={fill=orange!10, draw=orange!50!black},
    zgate/.style={fill=green!10, draw=green!50!black},
    phgate/.style={fill=red!10, draw=red!50!black}
}

\begin{document}

\title{A Zero-Added-Loss Multiplexing (ZALM) Source Simulation}

\author{Jerry Horgan$^{\dag}$, Alexander Nico-Katz$^{\ddag}$, Shelbi L. Jenkins$^{\S}$, Ashley N. Tittelbaugh$^{\S}$, Vivek Vasan$^{\dag}$, Rohan Bali$^{\S}$, Marco Ruffini$^{\dag}$, Boulat A. Bash$^{\S}$, and Daniel C. Kilper$^{\dag}$

\thanks{This work is supported by the Science Foundation Ireland grants 20/US/3708, 21/US-C2C/3750, and 13/RC/2077 P2 and National Science Foundation under Grant No. CNS-2107265.}
\thanks{$^{\dag}$J. Horgan, V. Vasan, M. Ruffini, and D. Kilper are with CONNECT Centre, Trinity College Dublin, Dublin, Ireland (e-mail: horganj3@tcd.ie; vasanv@tcd.ie; marco.ruffini@tcd.ie; dan.kilper@tcd.ie).}
\thanks{$^{\ddag}$A. Nico-Katz is with Trinity Quantum Alliance, Trinity College Dublin, Dublin, Ireland (e-mail: nicokata@tcd.ie; ).}
\thanks{$^{\S}$S. L. Jenkins, A. N. Tittlebaugh, and B. A. Bash is with the Electrical and Computer Engineering Department, University of Arizona, Tucson, AZ, USA (e-mail: boulat@arizona.edu).}
\vspace{-2em}}

\markboth{Journal of \LaTeX\ Class Files,~Vol.~14, No.~8, August~2021}%
{Shell \MakeLowercase{\textit{et al.}}: A Sample Article Using IEEEtran.cls for IEEE Journals}

\maketitle

\title{A Zero-Added-Loss Multiplexing (ZALM) Source Simulation}

\begin{abstract}
Zero-Added-Loss Multiplexing (ZALM) offers broadband, per-channel–heralded EPR pairs, with a rich parameter space that allows its performance to be tailored for specific applications. We present a modular ZALM simulator that demonstrates how design choices affect output rate and fidelity. Built in NetSquid with QSI controllers, it exposes 20+ tunable parameters, supports 'IDEAL' and 'REALISTIC' modes, and provides reusable components for  Spontaneous Parametric Down Conversion (SPDC) sources, interference, Dense Wavelength Division Multiplexing (DWDM) filtering, fiber delay, active polarization gates, detectors, and lossy fiber. Physics-based models capture Hong–Ou–Mandel (HOM) visibility, insertion loss, detector efficiency, gate errors, and attenuation. Using this tool, we map trade-offs among fidelity, link distance, and entangled pairs per use, and show how SPDC bandwidth and DWDM grid spacing steer performance. Using the default configuration settings, average fidelity remains constant at $\sim$0.83 but the ebit rate decreases from $\sim$0.0175 at the source to 0.0 at 50 km; narrowing the SPDC degeneracy bandwidth increases the ebit rate significantly without affecting fidelity. The simulator enables co-design of source, filtering, and feed-forward settings for specific quantum memories and integrates as a building block for end-to-end quantum-network studies.
\end{abstract}

\section{Introduction}\label{intro}

Robust entanglement distribution is a foundational requirement for the future quantum internet \cite{kimble2008quantum, wehner2018quantum}. While much research has focused on optimising the rate and fidelity of this entanglement distribution, the next step is integrating with quantum memories to form functional quantum memory networks. Adopting the principles of systems engineering, we employ a holistic, end-to-end analysis of the network architecture, following a modular approach consistent with prominent quantum network simulation frameworks \cite{Wu_2021, diadamo2020qunetsim}. This perspective integrates the performance of quantum entanglement sources, the fidelity degradation across quantum channels, and the efficiencies of quantum memories. While recent studies have successfully modelled entanglement generation between absorptive memories \cite{9951205} and simulated atomic ensembles \cite{zhou2023simulator}, these works primarily focus on the link-layer physics. We expand upon this body of knowledge by bridging the gap between component-level quantum optics and system-level network metrics. To address this, we model a novel quantum source specifically engineered for high-efficiency loading into quantum memories, demonstrating how optimizing the source-memory interface is critical for improving the performance of the entire quantum network.

Quantum entanglement sources come in many forms, with different natural wavelengths, emission efficiencies, and bandwidths \cite{eisaman2011invited}. Placing these sources into two main categories, we can look at their emission properties as either deterministic, such as the very precise wavelengths emitted by quantum dots, or probabilistic, whereby the specific emission wavelength of a generated photon pair is a stochastic variable distributed across a broad spectral bandwidth centred at a known degeneracy point. Although deterministic sources simplify network setups, the broadband capabilities of probabilistic sources allow for greater use of the available spectrum. This is enabled by adapting common classical network management techniques for routing and spectrum allocation \cite{jsac_rsa}.

To scale quantum networks beyond simple point-to-point links, quantum memories are an essential component. They provide the crucial capability to synchronise operations between distant nodes and to mitigate the inherently probabilistic nature of entanglement generation and swapping protocols. As such, memories form the cornerstone of quantum repeater architectures. The search for a practical physical implementation has led to significant research in solid-state systems. For instance, defects, or vacancies, in diamond are a practical and robust mechanism to provide quantum memories \cite{weber2010quantum}. Further work highlighted the attributes of the other group IV elemental defects in diamond, with Silicon showing considerable coherence times, greater than 1~s, and a more efficient spin-photon interface \cite{bradac2019quantum}. However, successfully interfacing with these memories is a significant challenge, with specific requirements on the incoming photons' properties, such as bandwidth and wavelength matching \cite{bersin2024telecom, PhysRevApplied.22.044013}.

The Zero-Added-Loss Multiplexing (ZALM) protocol provides for a broadband source with high memory loading efficiency \cite{Chen_ZALM, shapiro2024entanglement}. As there is currently no agreed implementation of the transmitter (Tx) component of this protocol and there are numerous configuration options of each of the transmitter subcomponents, we created a configurable simulation to determine the optimal transmitter architecture for maximising entanglement distribution rates and fidelity.

\section{The ZALM Source}\label{zalm}
ZALM, as described by \cite{Chen_ZALM} and \cite{shapiro2024entanglement}, is a broadband, deterministic source-in-the-middle architecture that is optimised for memory loading through the spin-photon interface. ZALM combines a pair of probabilistic SPDC sources into a quasi-deterministic quantum entanglement source where an entangled pair is emitted randomly but its timing and emission wavelength is known by the source and communicated to the receiver in advance of transmission. This information is used by the receiving node to efficiently convert the incoming photons for maximum compatibility with the memory loading process, hence there is `zero added loss' at the quantum receiver.

Whilst the emission rate of the ZALM source may be lower than other quantum entanglement sources, it has the advantage of only emitting entangled pairs that are capable of being loaded into a memory at the receiver. It classically heralds the receivers to expect a photon within a certain time frame allowing the unheralded (non entangled) photons to be discarded. 
\begin{figure}[ht]
	\centering	\includegraphics[width=\linewidth]{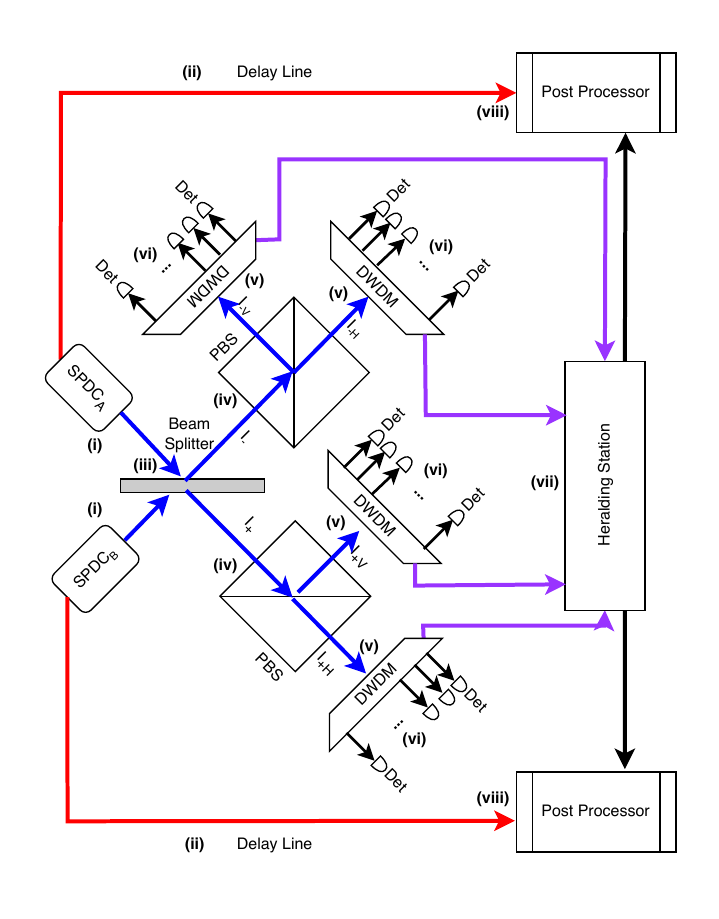}
	\caption{Schematic of the ZALM source components. Here $SPDC_{A|B}$ donates an SPDC process; $I_{+}$ and $I_{-}$ the positive or negative phase of an idler photon, further subscripted $I_{\pm{H|V}}$ to represent both phase and polarisation. PBS is a polarisation beam splitter; DWDM a DWDM filter with associate detectors (Det). }	\label{fig:zalm_source}
\end{figure}

The ZALM source consists of (i) two SPDC processes which are triggered simultaneously. These SPDC processes are configured to each produce a two-qubit entangled state (ebit). The qubits are physically photons\footnote{The terms `photons' and `qubits' are used interchangeably in the text. When discussing a photon, the authors generally discuss the physical properties associated with the photon and not the quantum state. When qubits are used, the authors focus on the quantum state of the system.} that are polarisation encoded. The photons produced by each SPDC process are labelled as signal and idler. The signal photons go directly to (ii) a delay line for later processing, whilst the idler photons are passed through (iii) a beam splitter (BS). 

The beam splitter, with some probability, will entangle the two idler photons, producing a four-qubit quantum state (which includes the two signal photons in the delay lines). The two idler photons are then emitted from the beam splitter where they pass through (iv) a polarising beam splitter (PBS), which separates them into their respective states. The PBSs are connected to (v) DWDM filters, each of which have as many (vi) detectors attached as channels available. Therefore, each detector has an associated central wavelength for the idler photon with a margin of error within $50\%$ of the DWDM channel width. The coincidence counts at the detectors are then heralded at the (vii) heralding station, which notifies the (viii) post-processor any corrections to be applied to the signal photons.

Knowing the central wavelength of the idlers, the approximate wavelength of the signal photons can be determined based on the principles of energy conservation used by the SPDC processes. Therefore, the smaller the DWDM channel bandwidth, the more accurate this approximation will be, improving the memory loading efficiency in the receiving node. This is the key feature of the ZALM source, and is what makes it a deterministic source.

The steps / components listed here are explained in greater detail in the following section.

\section{The ZALM Source Model}\label{model}
The ZALM source produces entanglement in multiple stages. Initially, entanglement is generated using SPDC processes, where a crystal is placed in a Sagnac configuration, see Fig.~\ref{fig:spdc_configurationA}. Two SPDC processes are used in the ZALM source and each creates, with some probability, a two-qubit quantum entangled state. These separate quantum states are then entangled, again with some probability, with each other to create a four-qubit state. These stages are now described in more detail.

\subsection{Spontaneous Parametric Down Conversion}
There are three types of SPDC processes: 0, I, and II. Each SPDC process is pumped by a laser, but the conservation of energy in each type acts differently, which affects the wavelength, emission angle, polarisation, and the bandwidth of the emitted photons. 

A type-0 SPDC process emits two photons with the same polarisation as the pump laser, creating a $|\Phi\rangle$ Bell state. Type-I SPDC processes emit photons at different angles. The emitted photons are both co-polarised, but they are orthogonal to the pump laser polarisation, again creating a $|\Phi\rangle$ Bell state. However, type-I processes are less efficient than type-0 and therefore are not used in the model. Type-II SPDC processes emit orthogonally polarised photons, and therefore emit a $|\Psi\rangle$ Bell state.
Note, that the wavelengths of each photon will be symmetric around the central wavelength, $\lambda$.

A degenerate SPDC process will create two photons, each with exactly half of the pump photon's energy, and will therefore be at the same wavelength. This is similar to a narrowband source and not desired in the ZALM source configuration. However, a non-degenerate SPDC process will unevenly distribute the pump photon's energy resulting in the signal and idler photons having different wavelengths. These wavelengths typically follow a Gaussian distribution across a known range and provide for a broadband source.

\subsubsection{Signal-Idler Separation}\label{sect:sig_idler}

The model allows for two distinct mechanisms to separate the signal and idler photons.

\begin{figure}[ht]
	\centering	\includegraphics[width=\linewidth]{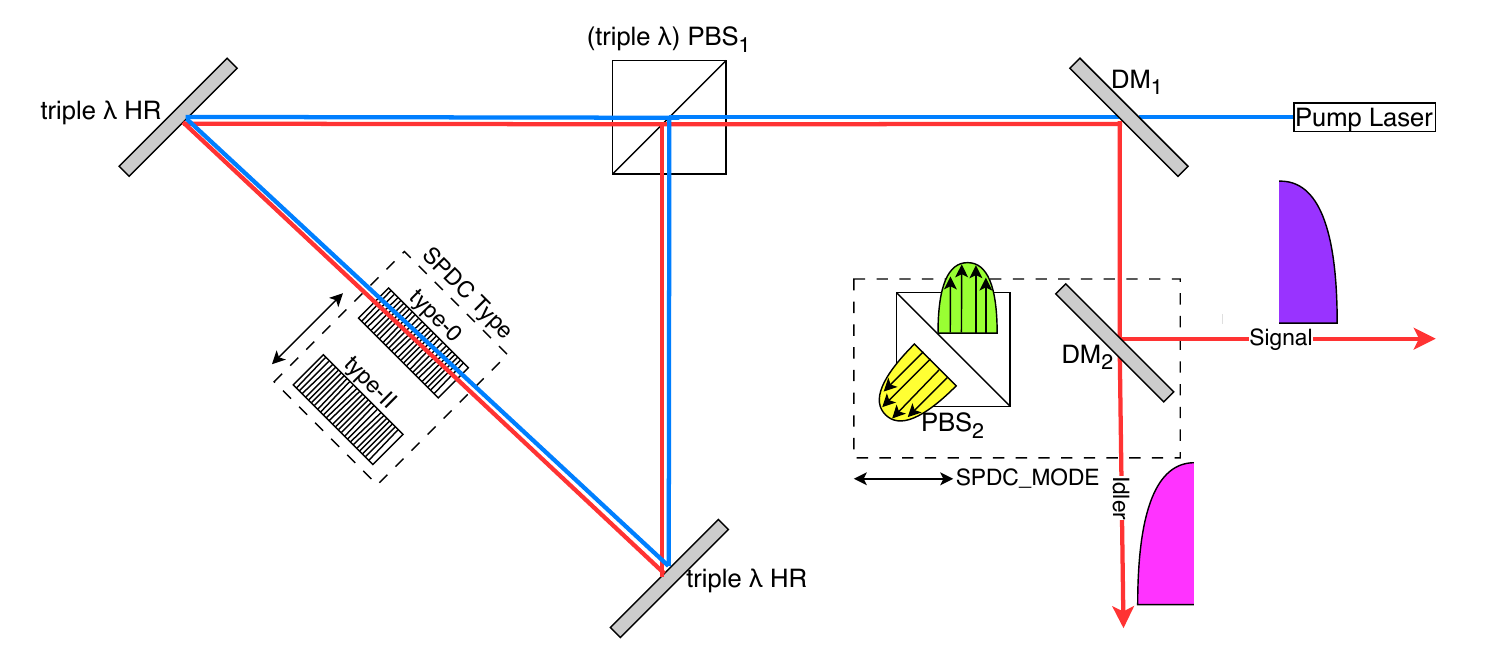}
	\caption{This schematic, which is based on Fig. 2 from the \cite{shapiro2024entanglement} paper shows two configuration options, SPDC\_Type and SPDC\_MODE, each indicated by the dashed boxes and double arrows. Here HR is a highly reflective mirror; $DM_{1}$ transmits the pump laser but is HR for the signal and idler photons. $PBS_{1}$ is a polarising beam splitter. $DM_{2}$ and $PBS_{2}$ are alternate configurations to separate the signal and idler photons.}	\label{fig:spdc_configurationA}
\end{figure}

\begin{enumerate}
    \item A polarising beam splitter can be configured to always have the H polarised photon be the signal and the V polarised photon be the idler, or vice versa. When the two SPDC processes are configured inversely to each other, i.e. with orthogonally polarised idlers, a $|\Psi\rangle$ state will always be produced.
    \item A Dichroic Mirror (DM) will separate photons based on a selected wavelength, which will either transmit or reflect the photon, see Fig.~\ref{fig:spdc_configurationA}.
\end{enumerate}

\subsection{Four-Qubit State Generation}

In the ZALM source, a 50:50 beam splitter is used to create entanglement between the two 2-qubit states, resulting in a four-qubit state. Whilst the Bell state generated at each SPDC process could be any of the four possible, it may not be known which encoded photon is the idler or the signal. Therefore, any combination of the four Bell states, $|\Phi^+\rangle$, $|\Psi^+\rangle$, $|\Psi^-\rangle$, or $|\Phi^-\rangle$ can be created at the beam splitter. 

\begin{table}[h]
\centering
\begin{tabular}{ l c c c r }
$|\Phi^+\rangle$ & = & $\frac{1}{\sqrt{2}}(|00\rangle + |11\rangle)$ & = &  $\frac{1}{\sqrt{2}}(|HH\rangle + |VV\rangle)$\\
$|\Psi^+\rangle$ & = & $\frac{1}{\sqrt{2}}(|01\rangle + |10\rangle)$ & = & $\frac{1}{\sqrt{2}}(|HV\rangle + |VH\rangle)$\\
$|\Psi^-\rangle$ & = & $\frac{1}{\sqrt{2}}(|01\rangle - |10\rangle)$ & = & $\frac{1}{\sqrt{2}}(|HV\rangle - |VH\rangle)$\\
$|\Phi^-\rangle$ & = & $\frac{1}{\sqrt{2}}(|00\rangle - |11\rangle)$ & = & $\frac{1}{\sqrt{2}}(|HH\rangle - |VV\rangle)$
    
\end{tabular}
\caption{The Bell states} \label{tab:bell_states}
\end{table}

\subsubsection{Beam Splitter}
The beam splitter has 2 input ports and 2 output ports. Each input port is connected to the idler output from an SPDC process, and each output port is connected to the input port of a polarising beam splitter. One of the output ports, the reflected port,  adds a 180\degree~phase shift. Photons will both exit at any one port if they are indistinguishable; otherwise there is a 50:50 chance that they will exit together through a single port or separate ports. The exit port(s) are selected randomly. 

As the SPDC process creates an H and V polarised pair of photons, any combination of H and V (HH, HV, VH, VV) can reach the beam splitter. This results in two scenarios when the photons are distinguishable (orthogonal polarisation) and two scenarios whereby they are indistinguishable.

Thus, statistically, in an ideal system, 50\% of the time the photons will be indistinguishable and will exit a single port and a further 50\% of the remaining photons will also exit a single port. Therefore, in an ideal system, 75\% of the photons will exit a single port. Furthermore, the beam splitter may introduce a phase shift, which in the global setting can be ignored, thereby enabling the creation of states $|\Phi^-\rangle$ and $|\Psi^-\rangle$.

\subsubsection{Hong–Ou–Mandel Visibility}\label{sect:hom}
No system is ideal, and a major source of decoherence is introduced by the Hong–Ou–Mandel (HOM) visibility of the photons that interact in the beam splitter \cite{HOMvisibility}. In the case of the ZALM source, the HOM visibility is concerned with how closely the two photons overlap spectrally, temporally, and in polarisation, with perfect overlapping in all three degrees of freedom producing completely indistinguishable photons. Anything less than perfect reduces the fidelity of the entanglement to the point that fidelity is zero if the photons are completely distinguishable. 

\subsubsection{Polarising Beam Splitter}
\begin{figure}[ht]
	\centering	\includegraphics[width=\linewidth]{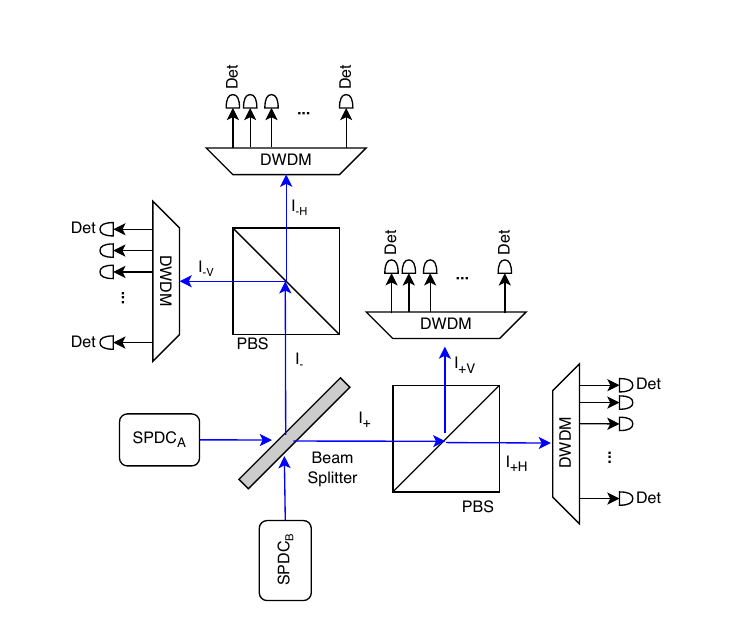}
	\caption{A subset of the components in Fig.~\ref{fig:zalm_source} to determine which Bell state is detected. $I_{\pm{H|V}}$ represents both phase and polarisation. If the polarisation is the same for both photons, the Bell state is $|\Phi^+\rangle$ or $|\Phi^-\rangle$ depending on the phase. If the polarisation is orthogonal, the Bell state will be $|\Psi^+\rangle$ or $|\Psi^-\rangle$ based on whether the phase is the same or different respectively.}	\label{fig:bell_states}
\end{figure}

The polarising beam splitter has a single input port\footnote{Vacuum noise is not modelled in the PBS.}, which may receive 0, 1, or 2 photons, and two output ports which may emit 0, 1, or 2 photons depending on how many photons were received, and their polarisation. The output ports from the polarising beam splitters are each connected to a dedicated DWDM filter. The output at the polarising beam splitters determine which of the four Bell states the quantum system is in (but not its fidelity). See Fig.~\ref{fig:bell_states} for details. 

If both photons are indistinguishable, they will exit the same port on both the beam splitter and the polarising beam splitter and will therefore create a $|\Phi^+\rangle$ or $|\Phi^-\rangle$ Bell state depending on whether they exited the $I_{+}$ or $I_{-}$ arm of the beam splitter. Photon-number-resolving detectors will be required to register the double click.

Otherwise, there is a 50\% chance that the photons will both exit the same port of the beam splitter but different ports of the polarising beam splitter. In this scenario a $|\Psi^+\rangle$ Bell state will be created as both photons will have the same phase. 

In the final scenario, both photons will exit separate ports on the beam splitter, where one photon will exit the H arm of one of the polarising beam splitters and the second photon the V arm of the other polarising beam splitter. This will produce a $|\Psi^-\rangle$ Bell state as there is a phase difference between both photons. See Table \ref{tab:bell_states} for a description of the polarisation-encoded Bell states.

\subsection{DWDM Filtering and Detection}

A DWDM filter is connected to each output of the polarising beam splitters. DWDM filters are assumed to have one detector per DWDM channel, regardless of channel bandwidth. 

\subsubsection{DWDM Channels}
Four identical DWDM filters are used to determine the approximate centre wavelength of the idler photons. These DWDM filters will typically use a grid spacing slightly larger than the emitted photon bandwidth and a guard-band to estimate the centre wavelength accurately. It is this specific feature that introduces the zero added loss characteristics of ZALM. 

\subsubsection{Heralding}

The detectors connected to the DWDM filters will determine which quantum state is expected based on the coincidence counts, see Fig.~\ref{fig:bell_states} for details. The detector click patterns, determined by phase and polarisation, and the central wavelength will be known. Any correcting information will be sent to the post processing unit and the timing and DWDM channel (wavelength) information will be sent to the receiver.

\subsection{Post Processing}

Based on the Bell state detected, some operations may be applied to the signal photons to transform them to the $|\Psi^-\rangle$ state\footnote{This action could be performed at the receiving side.}. A fibre delay loop is placed between the signal port on the SPDC process and Pockels cells which are tuned to implement the Pauli gates necessary. 

\subsubsection{Delay Line}

The processes within the ZALM source, from SPDC to heralding takes $\sim$20 ns. This equates to 4 m of a fibre loop delay line. This is used to provide sufficient time to herald the idler photons, and enables the operations (gates) to be applied to the photons being emitted into the network.

\subsubsection{\texorpdfstring{$|\Psi^-\rangle$ state (singlet)}{Psi- state (singlet)}}
The singlet state is the preferred choice for the quantum memory, as it can be loaded with very high (unit) fidelity using the augmented Duan-Kimble protocol \cite{PhysRevApplied.22.044013}. \\

Lastly, a simple circuit could be used to convert the polarisation encoded qubits into time-bin encoded qubits for more robust transmission over fibre \cite{shapiro2024entanglement}.

\section{Implementation / Results}
This ZALM source simulation is implemented in NetSquid \cite{coopmans2021netsquid}, and uses the QSI Controller and APIs \cite{QSI_Library_2022}. It can run standalone in NetSquid, or as part of a larger system within QSI. 

\begin{figure*}
	\centering	\includegraphics[width=\linewidth]{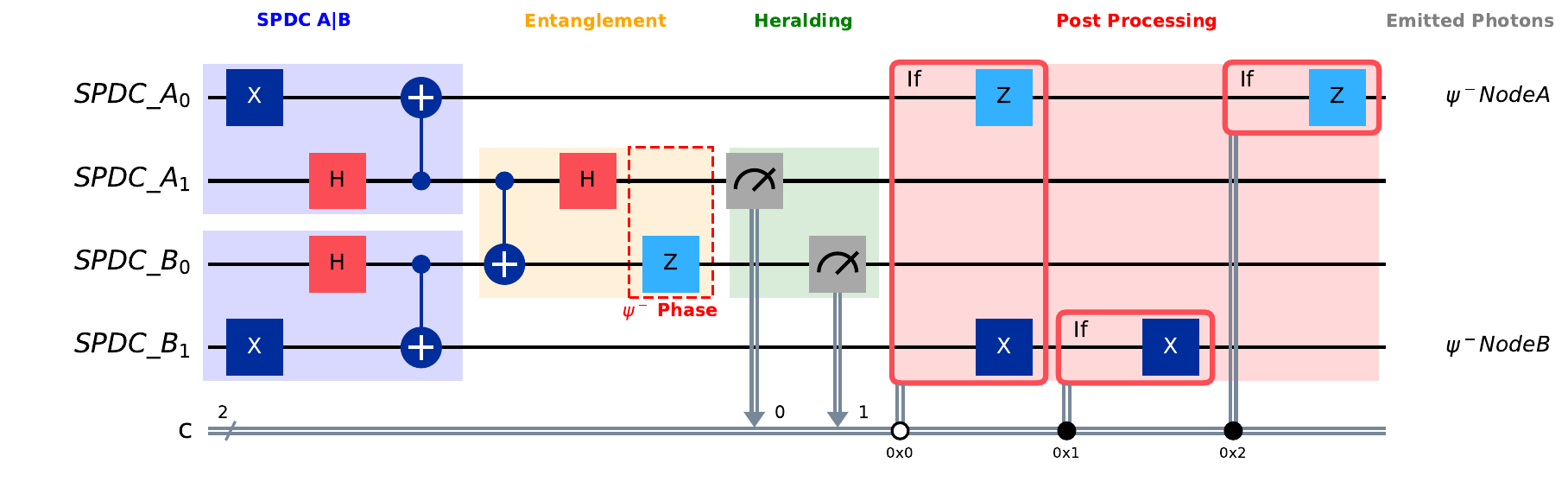}
	\caption{Circuit diagram of the quantum operations performed by the various QuantumPrograms.}	\label{fig:quantum_circuit}
\end{figure*}

The ZALM source is modelled on a table-top configuration where each component with conditional output paths is a distinct Node, i.e. the beam splitter is a separate node whereas the DWDM filters and the associated detectors are combined as the detectors have no output paths, rather they only register clicks. Each node has a quantum processor attached to it, which permits quantum operations (gates) to be applied to the qubits, Fig.~\ref{fig:quantum_circuit} details the quantum operations (gates/measurements) in a QisKit-style quantum circuit. Quantum noise models are applied to these quantum processors, as described in Section \ref{sec:config}. Furthermore, each node has a set of input and output ports, quantum channels, that are connected together via fibre of configurable length. Lastly, there is a classical information channel between the HeraldingStation and the PostProcessors.

A NodeProtocol runs on each node. The NodeProtocol is specific to the function of the node / component -- such as; BeamSplitterProtocol, FilterProtocol, PostProcessorProtocol, etc.~-- and will execute a QuantumProgram when direct manipulation of qubits is required, i.e. {SPDCInitQubits, SPDCProgram}, etc. Signals are used to communicate between the NodeProtocols. These Signals act as triggers to other NodeProtocols and are used to pass monitoring / status information. The schematic in Fig.~\ref{fig:protocol_structure} shows the ZALMProtocol, its sub-protocols, which can be mapped directly to the physical components in Fig.~\ref{fig:zalm_source}, and the signalling between them.

A FlyingQubit class, inherited from type Qubit, was created (custom\_qubits.py) to add additional properties, frequency and polarisation, to the default NetSquid qubit. A density matrix formalism is used to represent the quantum states. All qubits are stored in the quantum memory associated with each node. They are popped when an operation needs to be performed, including any gate or depolarising operations, and are then either placed back into memory or transmitted to the next port. The fibres between nodes have a fibre noise model applied.

A utils.py library provides supporting functions to the model, such as calculate\_visibility, determineBellState, generate\_flex\_dwdm\_grid, etc., system\_setup.py builds and connects the node configuration into a network, and config.py stores the default configuration parameters described in Section \ref{sec:config}. These parameters can be overridden in the config file. Additional implementation details are provided in Appendices B and C.

\begin{figure}[h]
\centering
\begin{tikzpicture}[node distance=0.2cm]

    
    \node[layer, fill=netsquidblue] (L1) {
        \textbf{SPDCProcessProtocol(NP)} \\
        \qfunc{ \quad SPDCInitQubits(QP)} \\
        \qfunc{ \quad SPDCProgram(QP)}
    };

    \node[layer, fill=netsquidblue, below=of L1] (L2) {
        \textbf{BeamSplitterProtocol(NP)} \\
        \qfunc{ \quad CNOTQubits(QP)} \\ 
        \cfunc{ \quad calculate\_visibility($\Delta_{f}$, $\Delta_{t}$)} \\
        \qfunc{ \quad PhaseFlipQubits(QP)}
    } ;

    \node[layer, fill=netsquidgreen, below=of L2] (L3) {
        \textbf{PBSProtocol(NP)} \\
        \textbf{FilterProtocol(NP)}
    };

    \node[layer, fill=netsquidpurple, below=of L3] (L4) {
        \textbf{HeraldingProtocol(NP)} \\
        \qfunc{ \quad BellMeasurementProgram(QP)}
    };

    \node[layer, fill=netsquidpurple, below=of L4] (L5) {
        \textbf{PostProcessorProtocol(NP)} \\
        \qfunc{ \quad PostProcessProgram(QP)}
    };

    \node[control, right=0.5cm of L1.north east, anchor=north west, minimum height=9.0cm] (Signal) {
        \textbf{Signal\\Protocol(NP)}\\
        \func{To keep Nodes in sync}
    };

    \foreach \i in {1,2,3,4,5} {
        \draw[<->, thick, orange!80, >=Stealth] (L\i.east) -- ($(Signal.west)!(L\i.east)!(Signal.west)$);
    }

    \begin{scope}[on background layer]
        \coordinate (TopPoint) at ($(Signal.north) + (0,15pt)$);
        \node[draw, dashed, gray, thick, rounded corners, inner sep=5pt, 
              fit=(L1) (L5) (Signal) (TopPoint),
              label={[anchor=north west, font=\sffamily\bfseries\Large]north west:ZALMProtocol}] (ZALM) {};
    \end{scope}

\end{tikzpicture}
\caption{Schematic showing the ZALMProtocol (ZALM Source) and its sub-protocols. QuantumPrograms are listed in blue, and major functions listed in red. (NP) and (QP) indicate a NetSquid NodeProtocol and QuantumProgram base class respectively.}\label{fig:protocol_structure}
\end{figure}

\subsection{Configuration options}\label{sec:config}

The model can be run in either of two modes by setting `SIM\_MODE' to either `IDEAL' or `REALISTIC'. The IDEAL mode introduces no losses or noise sources into the model, which is useful for tracking the quantum states through the system, whereas the REALISTIC mode operates with standardised values from real-world components. We now describe the default values for each configuration option in REALISTIC mode.

The SPDC process has 6 configurable parameters which specify the pump laser wavelength (nm), the degeneracy bandwidth of the SPDC processes (nm), the jitter between both processes (ps), the output photon bandwidth (GHz), the probability of a successful SPDC emission, and the signal/idler separation method:
\begin{center}
\fontsize{8pt}{8pt}\selectfont
\begin{tabular}{ l  r }
 SPDC\_PUMP\_WAVELENGTH\_NM & 775.0  \\ 
 SPDC\_DEGENERACY\_BANDWIDTH\_FWHM\_NM & 5.0  \\
 TEMPORAL\_JITTER\_STDEV\_PS & 0.0  \\
 EMISSION\_SUCCESS\_PROBABILITY & 0.95  \\
 PHOTON\_FWHM\_GHZ & 30.0 \\ 
 SPDC\_MODE & `DICHROIC'\\
  & `PBS' \\
\end{tabular}
\end{center}

The {BeamSplitter} has 2 configurable parameters which specify the minimum HOM visibility to be determined indistinguishable and the insertion loss associated with it:
\begin{center}
\fontsize{8pt}{8pt}\selectfont
\begin{tabular}{ l  r }
BEAMSPLITTER\_HOM\_THRESHOLD      &0.99\\
BEAMSPLITTER\_INSERTION\_LOSS\_DB &0.20
\end{tabular}
\end{center}

The {Polarising Beam Splitter} has 2 configurable parameters which specify the probability of a photon being emitted from the incorrect port and the insertion loss associated with it:
\begin{center}
\fontsize{8pt}{8pt}\selectfont
\begin{tabular}{ l  r }
PBS\_EXTINCTION\_RATIO      &0.001\\
PBS\_INSERTION\_LOSS\_DB &0.20\\
\end{tabular}
\end{center}

The {DWDM Filter} has 6 configurable parameters which specify which DWDM bands are enabled, the DWDM grid bandwidth (GHz), the size of any guard-bands, the insertion loss associated with it, and the filtering model in use:
\begin{center}
\fontsize{8pt}{8pt}\selectfont
\begin{tabular}{ l  r }
ENABLED\_BAND & [C, S, L]\\
GRID\_GRANULARITY\_GHZ & 100 \\
FILTER\_PASSBAND\_FRACTION & 0.8 \\
EFFECTIVE\_FILTER\_BANDWIDTH\_GHZ & 80 \\
DWDM\_FILTER\_INSERTION\_LOSS\_DB & 0.50\\
FILTER\_MODEL & `GAUSSIAN' \\
 & `BRICKWALL'
\end{tabular}
\end{center}

The {Detectors} have 2 configurable parameters, their detection efficiency and whether they are photon number resolving:
\begin{center}
\fontsize{8pt}{8pt}\selectfont
\begin{tabular}{ l  r }
DETECTOR\_EFFICIENCY &0.98 \\
DETECTOR\_TYPE &`STANDARD' \\
 &`PNR'
\end{tabular}
\end{center}

The QuantumProcessors have 4 configurable parameters, the gate error probability over a single gate, the gate error probability over two gates, the measurement dephasing probability, and the memory depolarization rate:
\begin{center}
\fontsize{8pt}{8pt}\selectfont
\begin{tabular}{ l  r }
GATE\_ERROR\_PROB\_SINGLE\_QUBIT & $1e^{-4}$ \\
GATE\_ERROR\_PROB\_TWO\_QUBIT & $1e^{-3}$ \\
MEASUREMENT\_DEPHASE\_PROB & $1e^{-3}$ \\ 
MEMORY\_DEPOLAR\_RATE & $1e^3$
\end{tabular}
\end{center}

The fibre INTERNODE\_LENGTH (km) is configurable between the heralding station and the measurement station to simulate a quantum channel:
\begin{center}
\fontsize{8pt}{8pt}\selectfont
\begin{tabular}{ l  r }
INTERNODE\_LENGTH &1$e^{-5}$
\end{tabular}
\end{center}

\subsection{Limitations}
The pump laser is not modelled in the SPDC process. The EMISSION\_SUCCESS\_PROBABILITY configuration option is used to set the efficiency level. It should also be noted that this source does not provide an ebit rate per second, but rather per use. Additionally, if the non-degeneracy bandwidth is very narrow and a dichroic mirror configuration is in use, the DM\_INSERTION\_LOSS\_DB and DM\_CROSSTALK\_PROBABILITY should be increased to either reflect the higher losses near the boundary wavelength, or else swap the component to a WSS or DWDM filter by increasing the insertion loss and lowering the cross talk probability. 

\subsection{Results}
Interacting with the simulator we execute the `qsi\_source\_fidelty.py' script to produce the following plots. This test script implements a measuring station 1 cm from the ZALM source. QSI activates `zalm\_source.py' and the QSI provided `fiber.py' modules. The fibre component is added twice, one for each connected Node. The fibre length, $L$, is set to 0 km as this is a test of the source. The following steps are repeated 10,000 times. The ZALM Source is then triggered via a `channel\_query' message from the QSI controller which returns the StateProp (density matrix, frequency, and polarisation) for each Signal qubit. The StateProp's of Signal 1 and 2 are separated and passed through their respective fiber channels, where any Kraus operators are applied to the StateProp, which should be none in this scenario. Finally, the resultant quantum state is checked against the expected $|\Psi^-\rangle$ state to determine fidelity, see Fig.~\ref{block:source}.
\begin{figure}[h]
\centering
\begin{tikzpicture}[node distance=1.0cm, every node/.style={draw, fill=white}]
    \node (init) {Initialize QSI Coordinator};
    \node (params) [below of=init] {Update Fiber $L=0$};
    \node (run) [below of=params] {Loop: $10^4$ Runs};
    \node (source) [below of=run] {Trigger ZALM Source};
    \node (fiber) [below of=source] {Apply Kraus Maps $\mathcal{E}_1, \mathcal{E}_2$};
    \node (fid) [below of=fiber] {Calculate Fidelity};
    
    \draw [->] (init) -- (params);
    \draw [->] (params) -- (run);
    \draw [->] (run) -- (source);
    \draw [->] (source) -- (fiber);
    \draw [->] (fiber) -- (fid);
    \draw [->] (fid.east) -- ++(1,0) |- (run.east); 
\end{tikzpicture}
\caption{qsi\_source.py block diagram}
\label{block:source}
\end{figure}

\begin{figure}[h]
	\centering	\includegraphics[width=\linewidth]{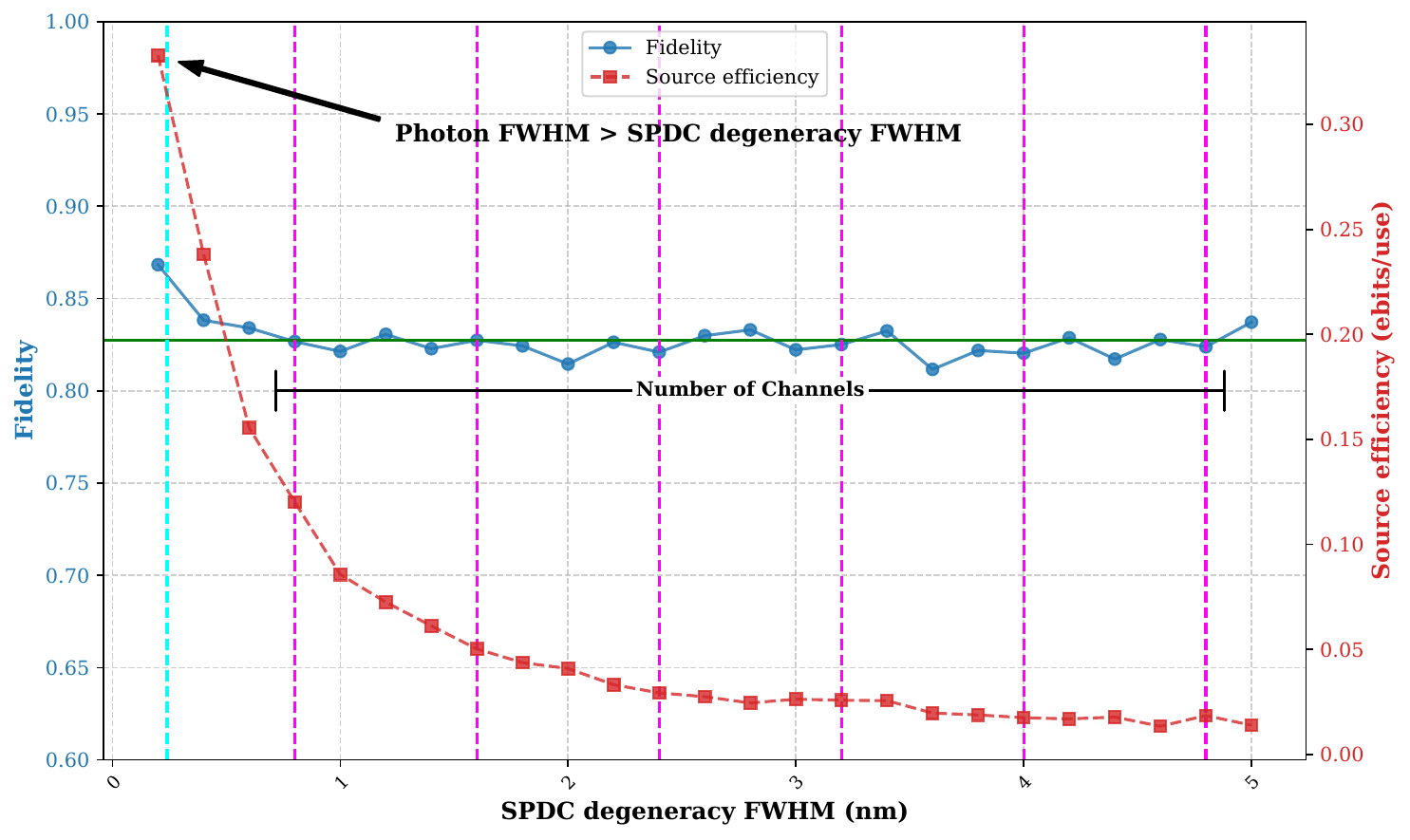}
	\caption{As SPDC degeneracy widens, source efficiency decreases. Fidelity also decreases but slower than efficiency. The cyan line represents the photon FWHM and the magenta lines each represent a 100 GHz DWDM channel. This plot shows SPDC FWHM degeneracy to 5 nm which covers 7 100 GHz DWDM channels.}	\label{fig:degeneracy}
\end{figure}

In Fig.~\ref{fig:degeneracy} we vary the SPDC degeneracy bandwidth from 0.2 nm to 5.0 nm in 0.2 nm steps. All other parameters are set to default. Two parameters of note in this plot are the photon FWHM bandwidth (30 GHz / 0.24 nm - cyan line) and the DWDM grid size (100 GHz / 0.8 nm spacing - magenta lines). As can be seen from the plot, source efficiency (in ebits per use) decreases as the SPDC degeneracy bandwidth increases. This is expected behaviour, as the chances of spectral overlap, and hence, HOM visibility, also decreases. Additionally, fidelity remains pretty constant at $\sim$0.83. It should also be noted that the first value on the plot is not possible in a realistic setup as the photon bandwidth is greater than that of the SPDC degeneracy bandwidth.

\begin{figure}[h]
	\centering	\includegraphics[width=\linewidth]{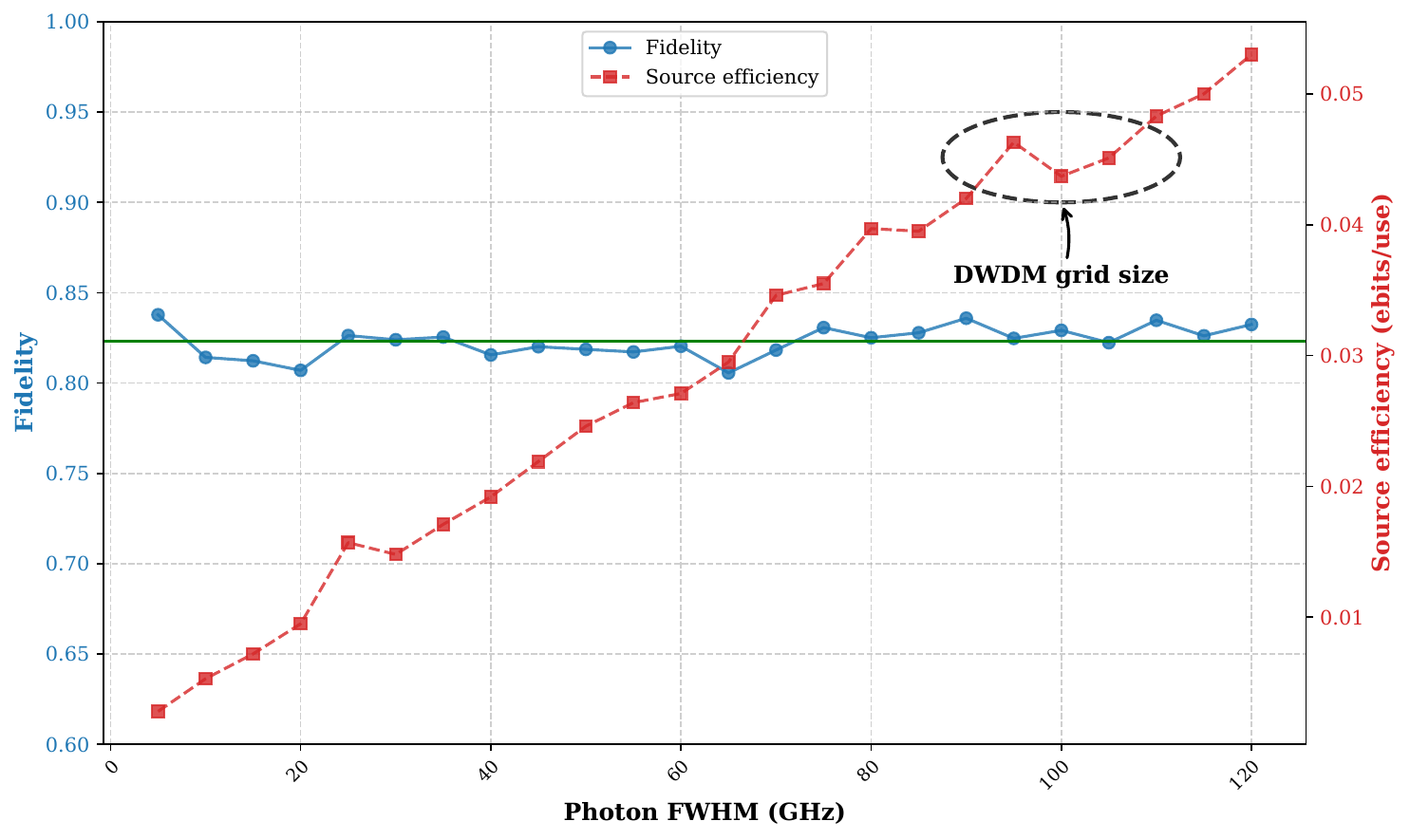}
	\caption{As the photon FWHM bandwidth increases so does the source efficiency. However, clipping occurs as the photon FWHM bandwidth approaches that of the DWDM grid spacing.}	\label{fig:photonfwhm}
\end{figure}

Increasing the photon FWHM bandwidth from 5 Ghz to 100 GHz in 5 GHz steps increases the source efficiency. This is expected behaviour, as the chance of spectral overlap, and hence, HOM visibility, also increases. Fidelity remains constant at $\sim$0.83. Additionally, as the photon bandwidth approaches the DWDM grid size, the source efficiency drops off due to spectral clipping, see Fig.~\ref{fig:photonfwhm}.

Figs. \ref{fig:degeneracy} and \ref{fig:photonfwhm} highlight the trade off between SPDC degeneracy bandwidth, photon bandwidth, and the source efficiency versus creating a multi-channel quantum source that can be spectrally sliced between several nodes. Consequently the more channels, $N$, are available the lower the overall source efficiency: $N \uparrow \implies \eta \downarrow$.

The potential of jitter between the two SPDC processes is modelled, increasing the standard deviation of both emissions from 0.0~ps to 20~ps in 0.5~ps increments. The relationship between the spectral bandwidth ($\Delta\nu$) and the temporal full-width at half-maximum ($FWHM_{t}$) is governed by the time-bandwidth product (TBP):
\begin{equation}
\Delta\nu \cdot FWHM_{t} \approx 0.441
\end{equation}

Given a configured spectral bandwidth of $\Delta\nu = 30$~GHz, the characteristic temporal duration of the photon is calculated as:
\begin{equation}
FWHM_{t} = \frac{0.441}{30 \times 10^9 \text{ Hz}} \approx 14.7 \text{ ps}
\end{equation}

To determine the visibility of the interference, we convert the physical $FWHM_{t}$ to the temporal standard deviation $\sigma_{t}$ using the standard Gaussian relationship:
\begin{equation}
\sigma_{t} = \frac{FWHM_{t}}{2\sqrt{2\ln{2}}} \approx \frac{14.7 \text{ ps}}{2.355} \approx 6.24 \text{ ps}
\end{equation}
where $2\sqrt{2\ln{2}} \approx 2.355$ is the constant relating the FWHM of a Gaussian distribution to its standard deviation.

The analytical visibility as a function of jitter is modelled as:
\begin{equation}
V(\sigma_{j}) = \frac{1}{\sqrt{1 + \left( \frac{\sigma_{j}}{\sigma_{t}} \right)^2}},
\end{equation}
where $\sigma_{j}$ is the RMS of the arrival time jitter.

\begin{figure}[h]
	\centering	\includegraphics[width=\linewidth]{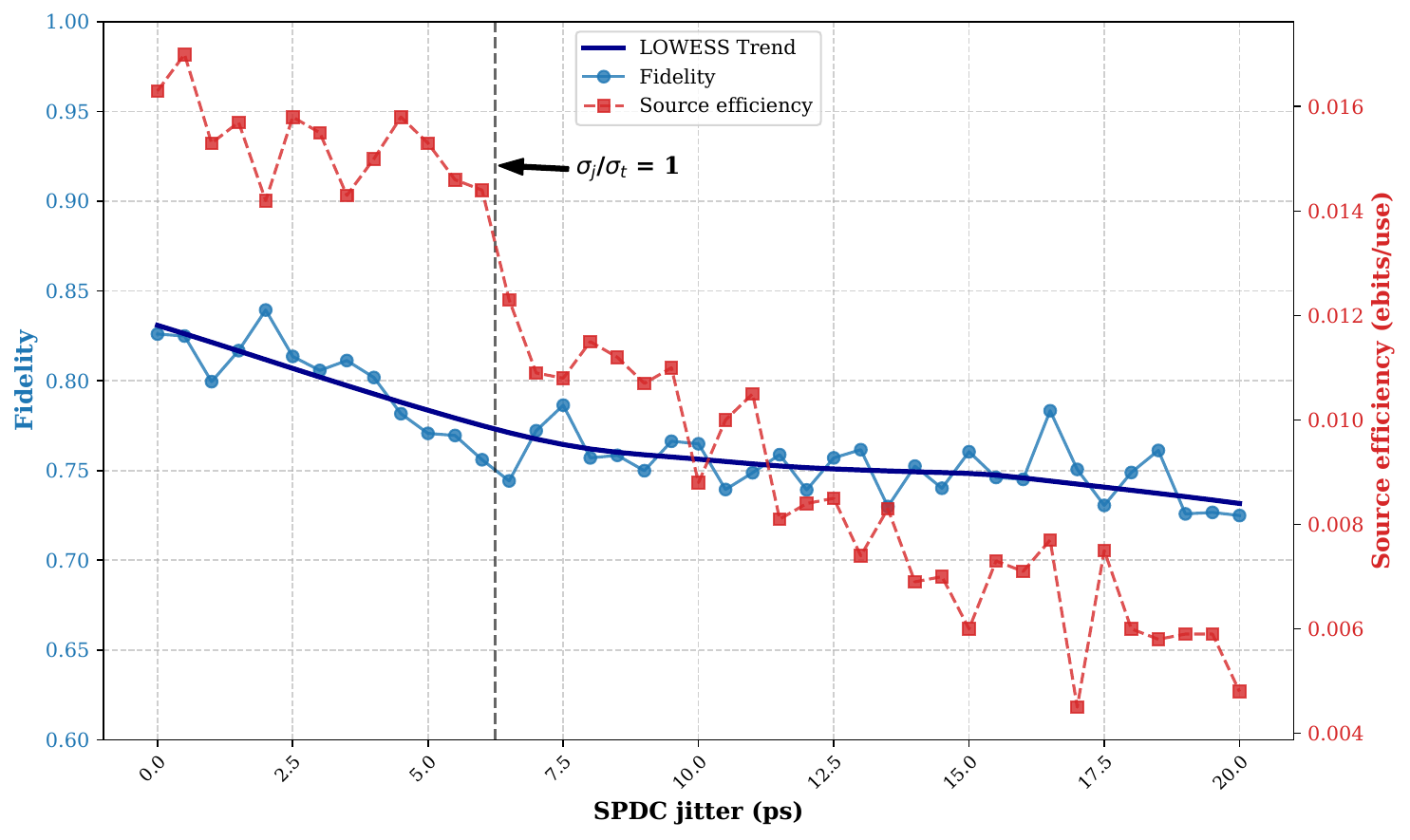}
	\caption{As SPDC jitter increases, entanglement fidelity and source efficiency decrease. $\sigma_{j}$ and $\sigma_{t}$ are the RMS of jitter and temporal standard deviation of the photon, respectively.}	\label{fig:spdcjitter}
\end{figure}

As expected, both fidelity and source efficiency decrease as jitter increases. The black line in Fig.~\ref{fig:spdcjitter} represents where the ratio of the RMS jitter, $\sigma_{j}$, over the temporal standard deviation of the photon, $\sigma_{t}$, is equal to 1.

\begin{figure}[h]
	\centering	\includegraphics[width=\linewidth]{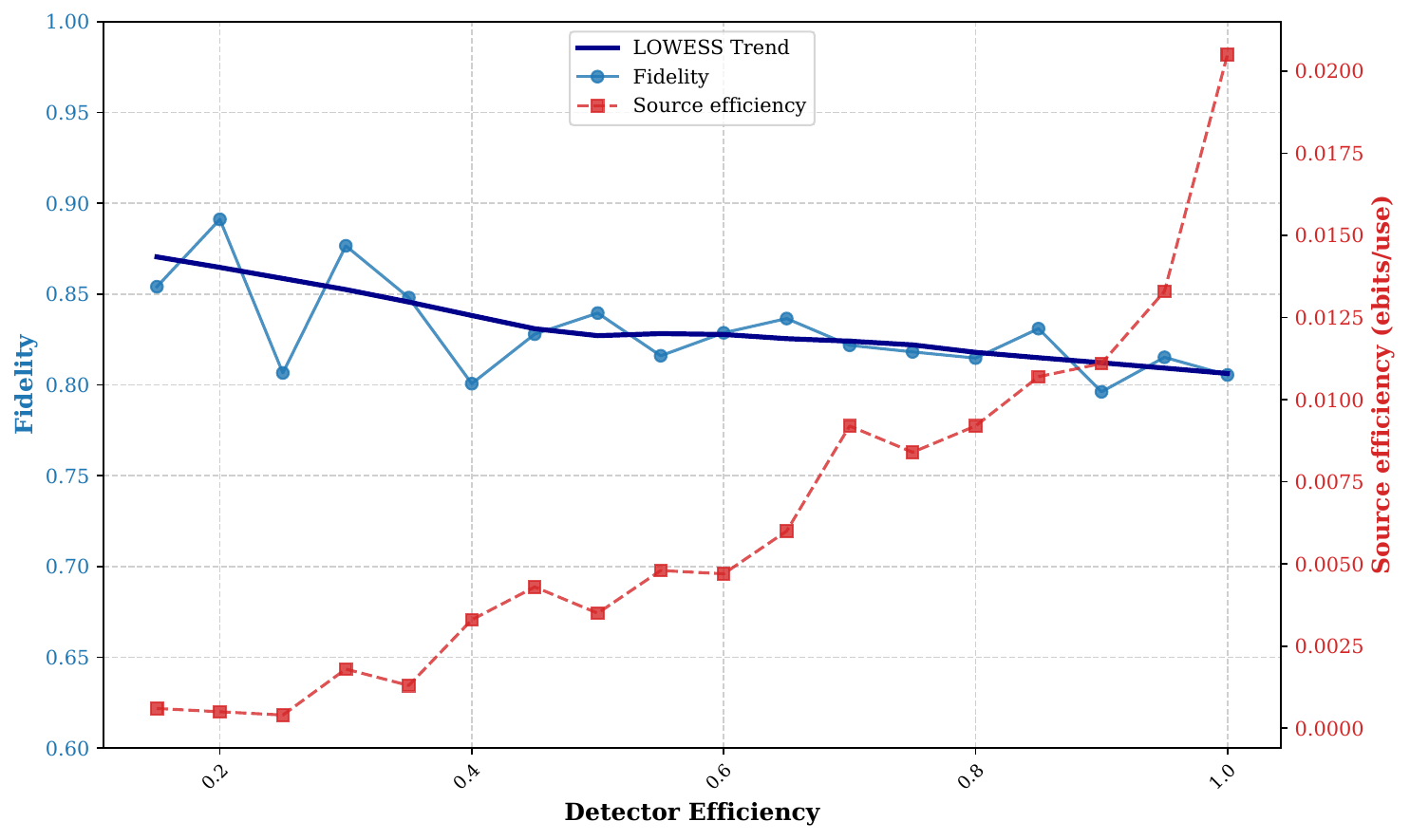}
	\caption{Fidelity and source efficiency vs.~detector efficiency. }	\label{fig:deteff}
\end{figure}

In Fig.~\ref{fig:deteff} we explore the impact of the efficiency of the detectors at the heralding station. As expected, improving it from 0.15 to 1.0 in 0.5 increments increases the source efficiency. Mean fidelity remains $\sim$ 0.83, however, fidelity appears to decrease as detector efficiency improves. We assume this due to only the higher fidelity entangled photon pairs being successfully detected when detector efficiency is low. 

\begin{figure}[h]
	\centering	\includegraphics[width=\linewidth]{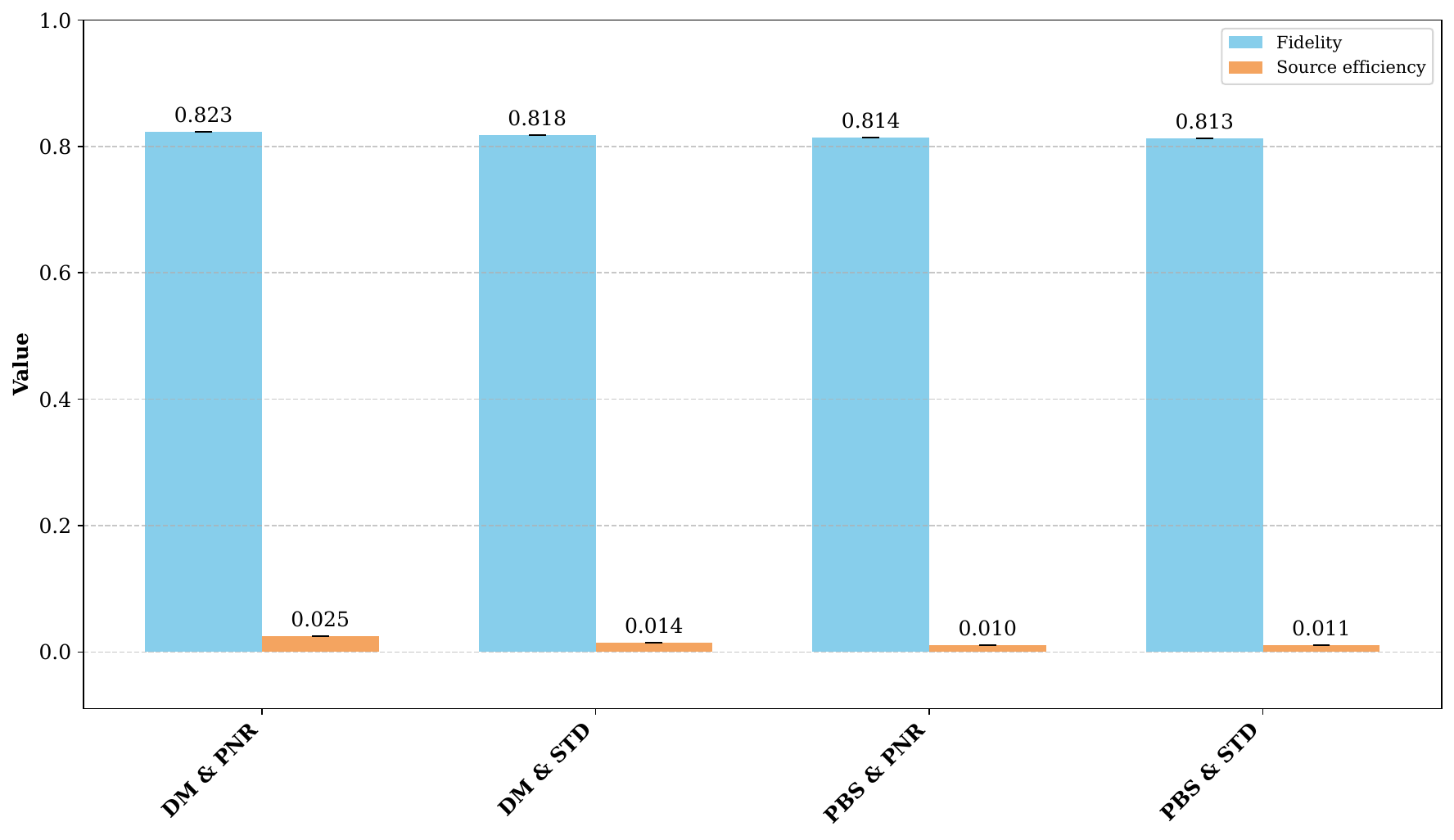}
	\caption{Mixed SPDC Modes and Detector Types. DM and PBS indicates a dichroic mirror or polarising beam splitter signal-idler separation configuration was used respectively. PNR and STD indicates that photon number resolving or standard detectors respectively were used. }	\label{fig:mode_det}
\end{figure}

We explore the performance metrics by varying the SPDC mode and the detector types between dichroic mirror (DM) or polarising beam splitter (PBS) and photon number resolving (PNR) or standard (STD) respectively. Standard detectors can only detect $|\Psi\rangle$ Bell states, whereas photon number resolving, by which we mean 0,1 or >1, detectors can detect both $|\Psi\rangle$ and $|\Phi\rangle$ Bell states. This can potentially double the source efficiency. Both the PBS and DM modes can produce all four Bell states, however, as discussed in Section \ref{sect:sig_idler} there is a mechanism to ensure only $|\Psi\rangle$ Bell states are produced using the PBS mode by setting each SPDC process' output idlers orthogonally polarised to each other. However, this PBS configuration does not appear to provide any advantage over using a DM mode. Using the DM mode, the SPDC degeneracy can be effectively split, by ensuring that the idler photons are situated in the longer-wavelength region of the spectrum relative to the degeneracy point, and thus reducing the spectral difference between idlers. As can be seen in Fig.~\ref{fig:mode_det} the DM-PNR combination provides the highest source efficiency, followed by the DM-STD combination. Fidelity varies minimally between each combination. 

\begin{figure}[ht]
	\centering	\includegraphics[width=\linewidth]{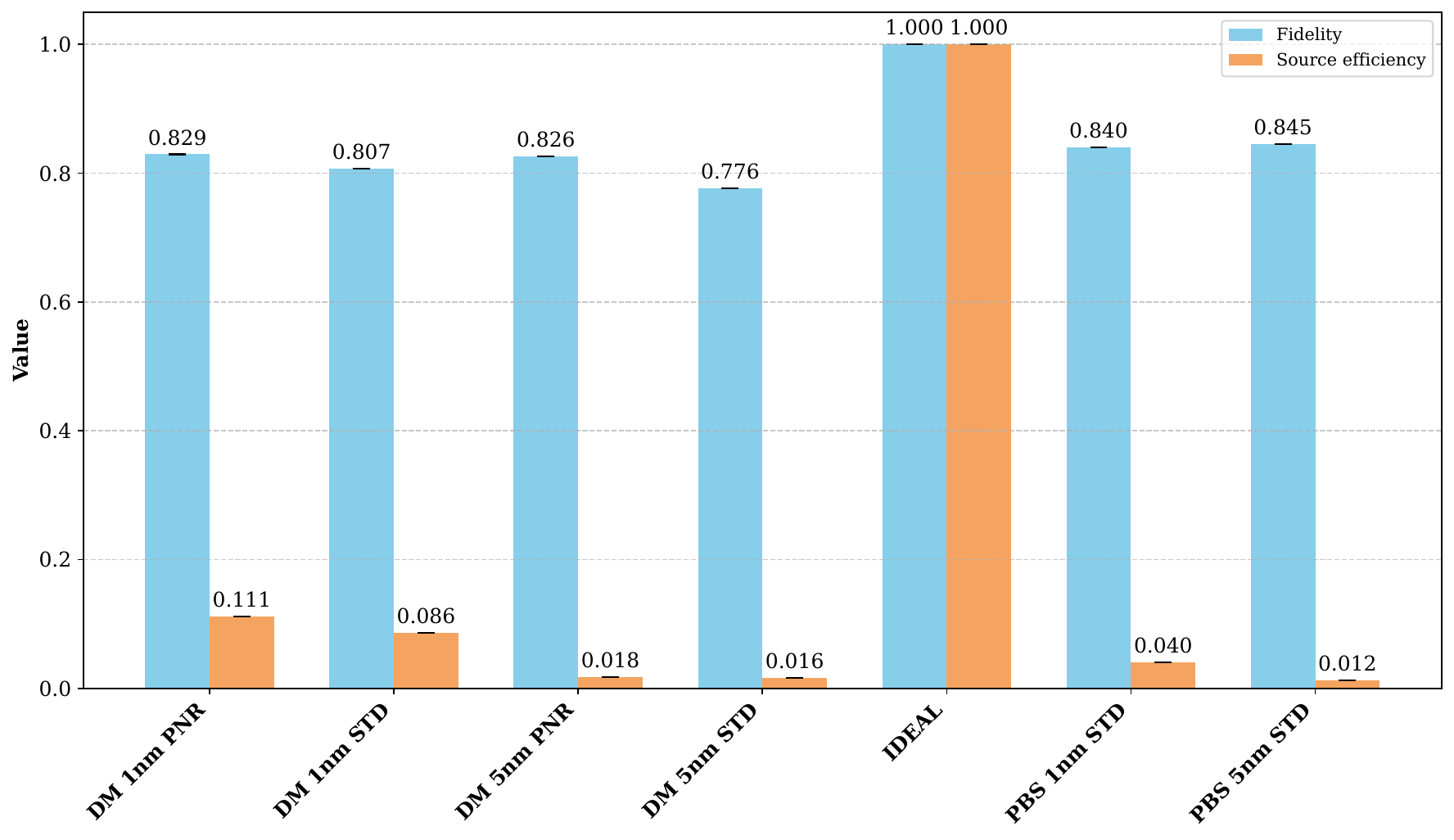}
	\caption{Sample of results from different parameter configurations. DM and PBS indicates dichroic mirror or polarising beam splitter. `x'~nm indicates the SPDC degeneracy bandwidth. PNR and STD indicates that photon number resolving or standard detectors respectively were used.  }	\label{fig:source_values}
\end{figure}

Fig.~\ref{fig:source_values} demonstrates the configurability of the ZALM source model. It shows the average fidelity and number of emitted qubits at the source. Here we demonstrate an IDEAL system configuration, along with SPDC non-degeneracies of 1~nm and 5~nm, using dichroic mirrors or polarising beam splitters for idler separation, and standard or photon number resolving detectors. Whilst there is a reduction and some variation in the entanglement fidelity in comparison to the ideal source, there is a more pronounced impact on the source efficiency. 

\begin{figure}[ht]
	\centering	\includegraphics[width=\linewidth]{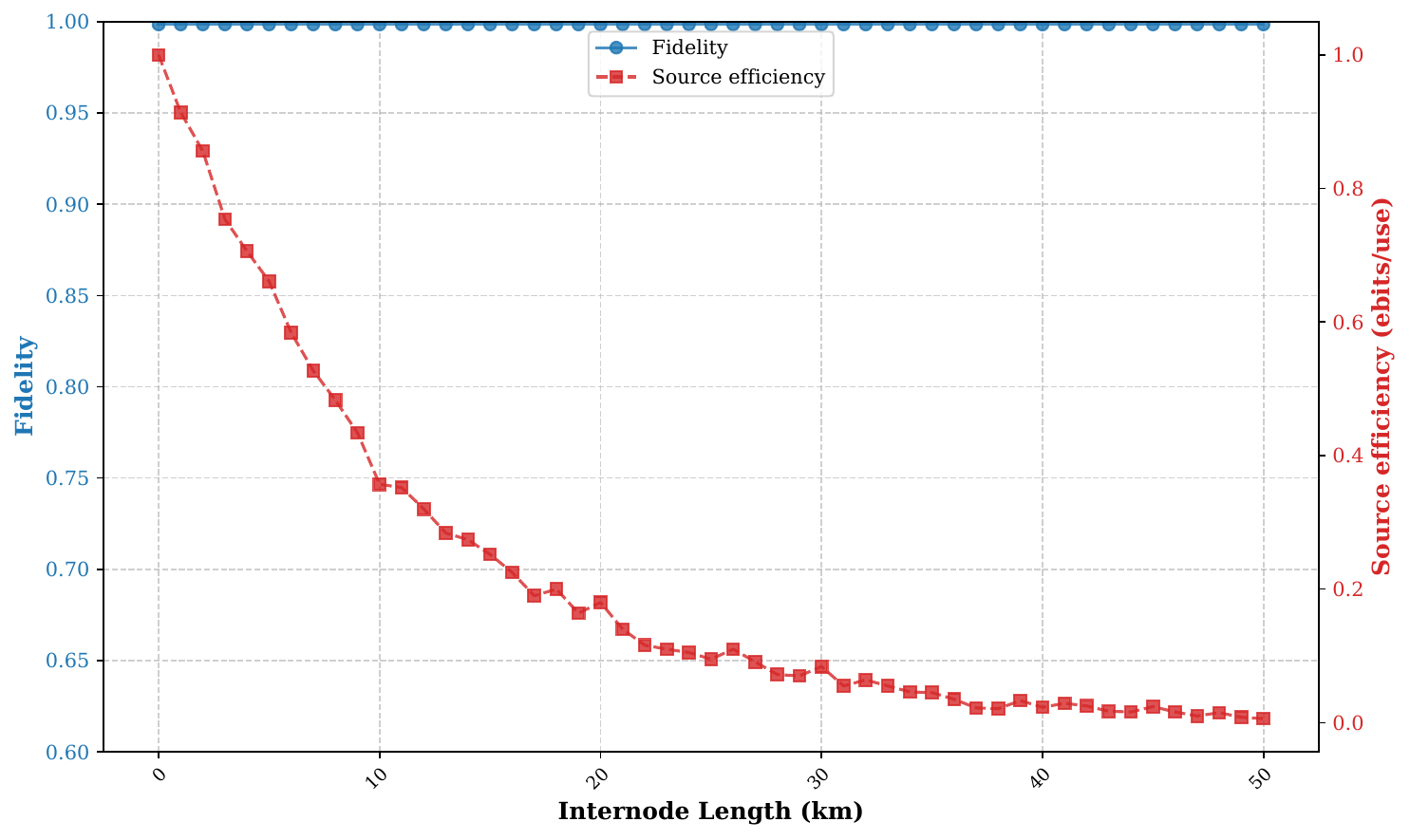}
	\caption{Fidelity and source efficiency vs.~distance using IDEAL ZALM source settings and fibre loss.}	\label{fig:fid_vs_dist_ideal}
\end{figure}

We now interact with the simulator using another testing script, qsi\_distance\_vs\_fidelity.py. Running the ZALM source in IDEAL mode and plotting the average fidelity and ebit rate vs distance, the initial fidelity is $\sim$1.0, which does not degrade over distance. However, the ebit rate drops inline with fibre losses, see Fig.~\ref{fig:fid_vs_dist_ideal}. It should be noted that ZALM uses a source in the middle model configuration and therefore fibre distances are effectively doubled\footnote{All fibre distances will be symmetric in the results shown.}. 

\begin{figure}[ht]
	\centering	\includegraphics[width=\linewidth]{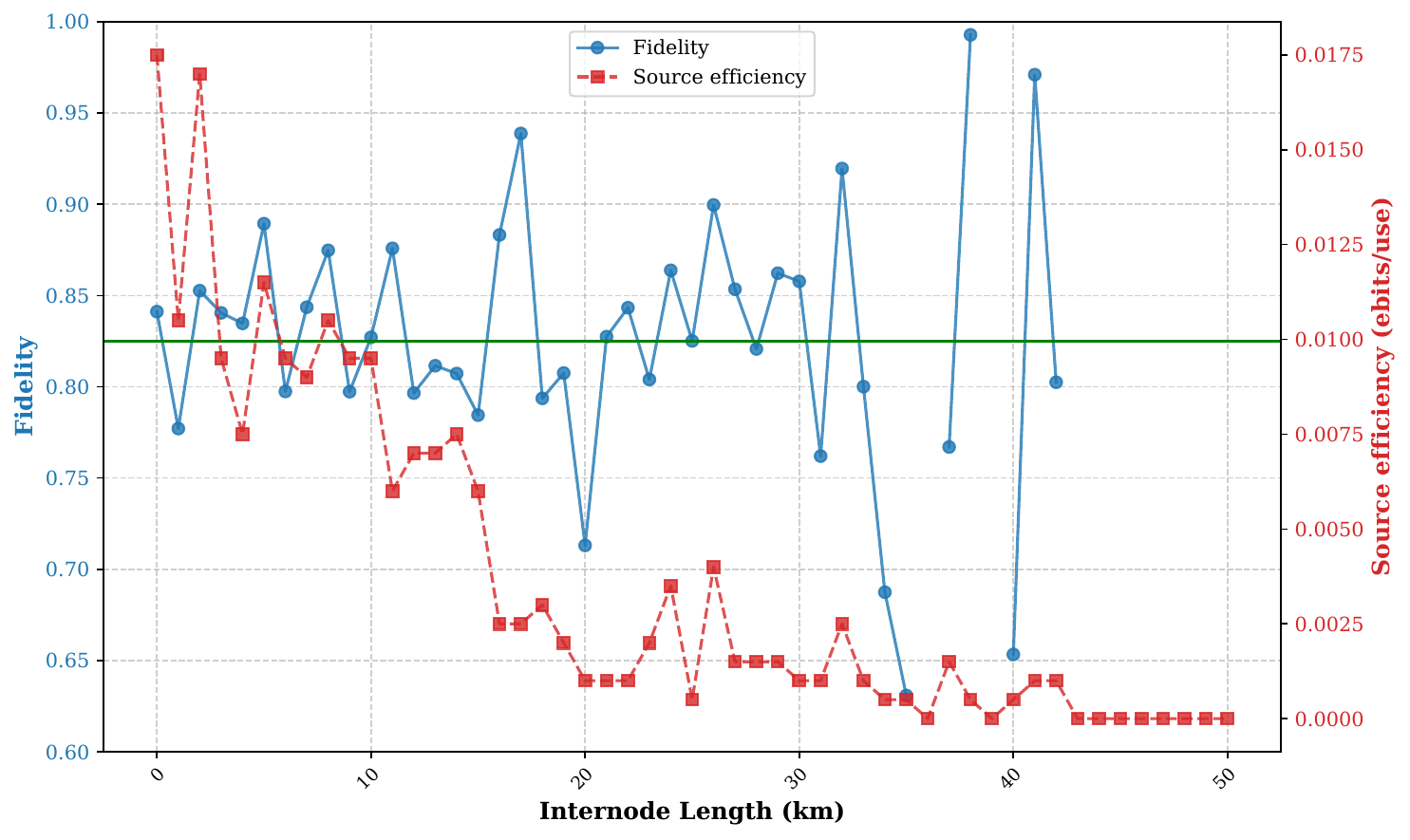}
	\caption{Fidelity and source efficiency vs.~distance using default ZALM source settings. Poor source efficiency, often 0.0000, after 30~km results in highly dynamic fidelity readings. Average fidelity over 50~km is marked by the green line.}	\label{fig:fid_vs_dist_default}
\end{figure}

Using the default configuration parameters, the fidelity and ebit rate of the distributed entanglement is plotted against distance from 0 to 50~km, see Fig.~\ref{fig:fid_vs_dist_default}. 

It is clear that the ebit rate dramatically falls off, but the average fidelity remains high enough to be purified, which suggests that network performance will be limited by the latency of classical signaling for purification rather than the intrinsic quality of the raw entanglement.

\begin{figure}[ht]
	\centering	\includegraphics[width=\linewidth]{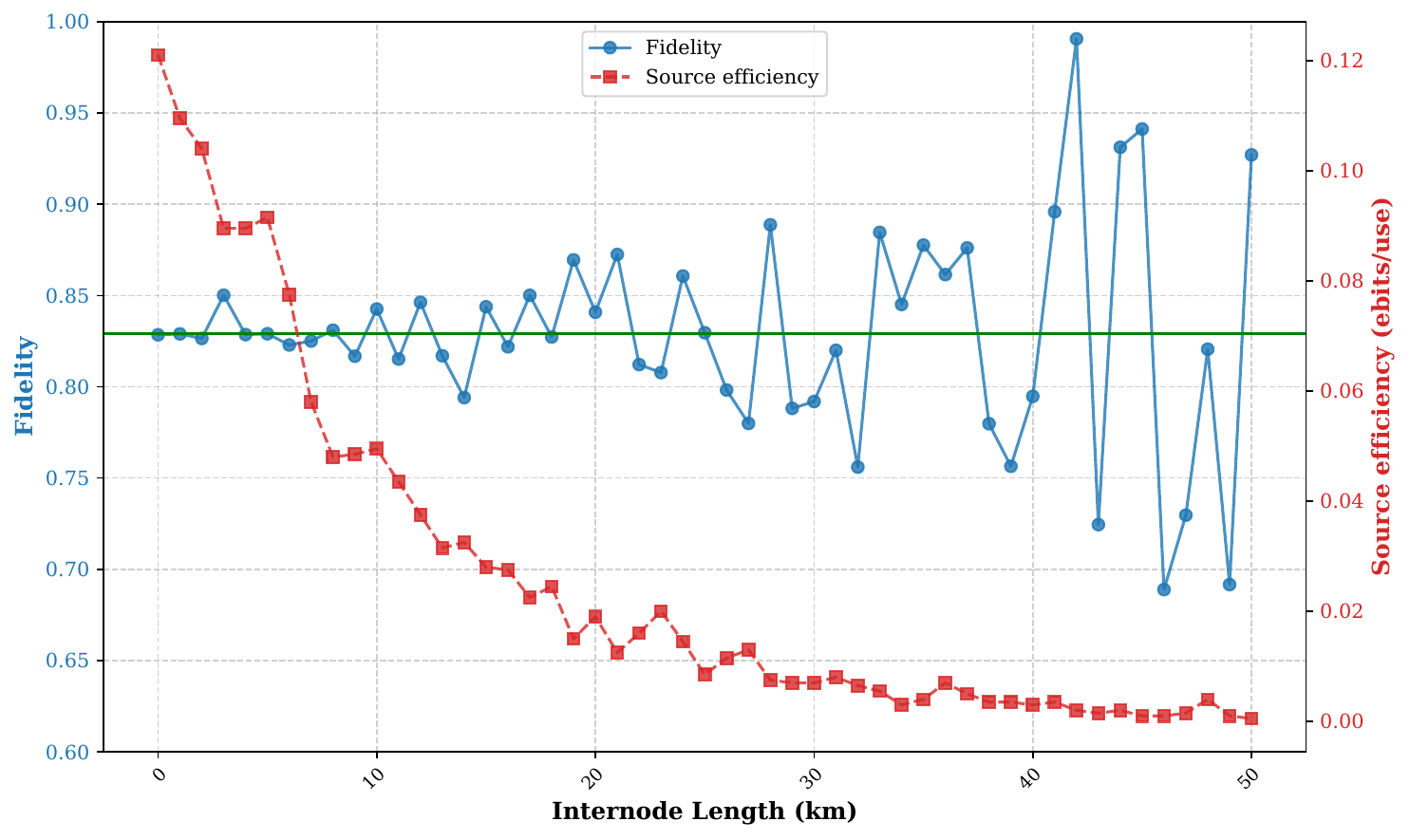}
	\caption{Fidelity and source efficiency vs.~distance using default ZALM source settings except SPDC degeneracy FWHM bandwidth of 1~nm and PNR detectors. Average fidelity over 50~km is marked by the green line.}	\label{fig:fid_vs_dist_1}
\end{figure}

Changing the SPDC\_DEGENERACY\_BANDWIDTH\_FW-HM\_NM to 1.0 and the DETECTOR\_TYPE to PNR improves the ebit rate considerably, whilst the average fidelity remains roughly the same, see Fig.~\ref{fig:fid_vs_dist_1}.

\section{Conclusions}\label{sec5}

We have developed a modular simulation of a ZALM-type entanglement source. The model is highly configurable with over 20 key physical parameters that can be adjusted. Our simulation serves as a tool to co-design and optimise the source for maximal network efficiency. Furthermore, it provides a modular building block within the QSI library where larger quantum simulations can be created, allowing full end-to-end performance analysis of various component configurations to identify optimal operating modes for specific network architectures.

\section{Future Work}\label{sec6}

Using the ZALM Source Simulator further parameters can be adjusted to determine the best ratio between SPDC degeneracy bandwidth, the photon bandwidth, and the DWDM grid spacing to provide a high-fidelity, high-rate,  multi-channel quantum entanglement source. 

In the future work, we will adjust the exiting SPDC parameters to implement new advances in ZALM source research, such as the phase-matched spectral islands architecture \cite{zalm-islands}.


\section*{Acknowledgments}
This work is supported by the Science Foundation Ireland grants 20/US/3708, 21/US-C2C/3750, and 13/RC/2077 P2 and National Science Foundation under Grant No. CNS-2107265.

\section*{Data Availability}
The data is available from the authors upon reasonable request. Code is freely available on GitHub \cite{Horgan_ZALM_Source_Simulation_2025}, under the GPLv3 license.

\section*{Conflict of interest}
The authors declare no potential conflict of interests.

\bibliographystyle{IEEEtran}
\bibliography{main}



\appendix

\section{SPDC Configurations\label{appa}}
\vspace*{12pt}
As mentioned in Section \ref{sect:sig_idler} either a type-0 or a type-II SPDC process can be used for the ZALM source. As the type-0 encodes the photons with the same polarisation a dual-wavelength half wave plate needs to be added to the setup to flip the polarisation of either the signal or the idler photons. This will then allow for signal-idler separation via wavelength or polarisation.

\section{ZALm Source Protocols\label{appb}}
\vspace*{12pt}
The Protocols (zalm\_protocols.py) provide the logic for each stage of the simulation. They are all implemented as NetSquid NodeProtcols. Some of the protocols execute NetSquid QuantumPrograms or equations to calculate noise or loss probabilities, but they primarily implement flow control, send and receive qubits, and track the quantum state. Noise is applied as a depolarising channel which is implemented in NetSquid as a Pauli channel. This noise channel can be explicitly called using the NetSquid qubitapi.depolarize(qubit, prob)\footnote{``Note that the probability that the quantum state of a qubit is actually changed due to application of a Pauli operator is 0.75 * prob." \cite{NetSquidAPI}} function, which is used to apply the HOM visibility, $V$. 

\begin{align*}
qubitapi.depolarize(qubit, prob=1 - V)
\end{align*}

Otherwise, noise is applied using an ErrorModel. The QuantumProcessor, where all gate and measurement operations are performed - see Fig.~\ref{fig:quantum_circuit}, uses two models, i) a DepolarNoiseModel - applied to all gate operations, and ii) a DephaseNoiseModel - applied to measurement operations. Both models take a probability argument. Additionally, all the modelled physical components within the ZALM source, see Fig.~\ref{fig:zalm_source}, are connected by 1 cm of fibre. Again, two ErrorModels are applied to any fibre, i) a DepolarNoiseModel, and ii) a FibreLossModel. The FibreLossModel applies an attenuation loss, $\alpha$ ($dB/km$). In this simulation all insertion losses are calculated as a probability, using numpy.random.rand(). As we are using the density matrix (DM) formalism the qubit is traced out. This can be applied as:
\begin{align*}
qubitapi.discard(qubit)
\end{align*}
 Listings 1-6 provide snippets of the simulation code for the BeamSplitterProtocol. A schematic of the ZALM Protocol and its associated sub protocols is shown in Fig.~\ref{fig:protocol_structure}.

\section{Beam Splitter Protocol\label{appc}}
\vspace*{12pt}
The 50:50 beam splitter is where the most quantum noise is introduced in terms of creating a single four qubit quantum state. This is primarily driven by the HOM Visibility, $V_{HOM}$ as described in Section \ref{sect:hom}. In our implementation this is determined through calculating the difference in photon frequencies, $\Delta{f}$, and their arrival time at the beam splitter, $\Delta{t}$,

\begin{equation}
\label{eq:total_visibility}
V_{HOM}(\Delta f, \Delta t) = V_{\nu}(\Delta f) \cdot V_{\tau}(\Delta t),
\end{equation}

where the spectral and temporal visibilities are modelled as Gaussian functions:

\begin{equation}
V_{\nu}(\Delta f) = \exp \left( -\frac{\Delta f^2}{2\sigma_f^2} \right), \quad 
V_{\tau}(\Delta t) = \exp \left( -\frac{\Delta t^2}{2\sigma_t^2} \right).
\end{equation}

The standard deviations $\sigma_f$ and $\sigma_t$ are related to the input full-width at half-maximum (FWHM) parameters, $\Delta \nu$ and $\Delta \tau$, by:

\begin{equation}
\sigma_f = \frac{\Delta \nu}{2\sqrt{2\ln 2}}, \quad \sigma_t = \frac{\Delta \tau}{2\sqrt{2\ln 2}}.
\end{equation}

In this model, the temporal FWHM duration $\Delta \tau$ is derived from the spectral bandwidth $\Delta \nu$ (in GHz) as:
\begin{equation}
\Delta \tau = \frac{1000}{\Delta \nu} \, [\text{ps}].
\end{equation}

The entangling QuantumProgram at the beam splitter is shown in Listing 1. The resultant entanglement would be of very high fidelity at this point, however the HOM visibility is then calculated and noise is applied.

\begin{lstlisting}[caption={Entangle the Qubits at the beam splitter},label=CNOTqubits, basicstyle=\fontsize{8}{10}\selectfont\ttfamily]
class CNOTQubits(QuantumProgram):
    default_num_qubits = -1

    def program(self):
        q1, q2 = self.get_qubit_indices(2)
        self.apply(instr.INSTR_CNOT, [q1, q2])
        self.apply(instr.INSTR_H, q1)

        yield self.run(parallel=False)
\end{lstlisting}

The depolarising probability is calculated as $1-V_{HOM}$, which is then applied to each qubit in the beam splitter, as shown in Listing 2. 

\begin{lstlisting}[caption={Qubit Depolarisation due to HOM Visibility},label=HOMdepolar, basicstyle=\fontsize{8}{10}\selectfont\ttfamily]
depolarizing_prob = 1 - visibility
if depolarizing_prob > 1e-9: 
    noisy1, noisy2 = self.node.qmemory.pop([0, 1])
    qapi.depolarize(noisy1, prob=depolarizing_prob)
    qapi.depolarize(noisy2, prob=depolarizing_prob)
    self.node.qmemory.put([noisy1, noisy2])
\end{lstlisting}

Listing 3 details the PhaseFlipQubits QuantumProgram. It takes a single parameter, flipQubits, which contains a list of 1 or 2 qubits that require a phase flip operation to be applied.

\begin{lstlisting}[caption={Phase Flip a Qubit},label=PhaseFlip, basicstyle=\fontsize{8}{10}\selectfont\ttfamily]
class PhaseFlipQubits(QuantumProgram):
    default_num_qubits = 2

    def program(self, flipQubits):
        for qubit in flipQubits:
            self.apply(instr.INSTR_Z, qubit)

        yield self.run()
\end{lstlisting}

Insertion, or any other efficiency, loss is demonstrated in Listing 4. Here the BEAMSPLITTER\_INSERTION\_LOSS\_DB is defined in decibels and therefore converted to linear form before numpy.random.rand() is called for each input qubit. If either random value is below the linear insertion loss threshold both qubits are discarded (lost).

\begin{lstlisting}[caption={Beamsplitter Insertion Loss},label=BSil, basicstyle=\fontsize{8}{10}\selectfont\ttfamily]
loss_prob = // 
1 - 10**(-config.BEAMSPLITTER_INSERTION_LOSS_DB / 10)
if np.random.rand() < loss_prob or np.random.rand() // 
< loss_prob: # 2 qubits
    qapi.discard(qa[0])
    qapi.discard(qb[0])
\end{lstlisting}

Listing 5 shows the code to generate the $\Delta_{f}$ and $\Delta_{t}$ values for the calculate\_visibility function.

\begin{figure*}[h]
\begin{lstlisting}[caption={Calculate the SPDC jitter and derive the HOM visibility, $V_{HOM}$, of the qubits recieved at the beam splitter.},label=HOMvisibility1, basicstyle=\fontsize{8}{10}\selectfont\ttfamily]
# Applying any post entanglement depolarisation - may remove entanglement
# Calculating HOM Visibility
delta_f_ghz = abs(q1.wavelength - q2.wavelength)/1e9 # Convert from Hz to GHz
# Calculate the jitter differential between both incoming photons.
jitter_sigma_ps = config.TEMPORAL_JITTER_STDEV_PS
jitter_a = np.random.normal(0, jitter_sigma_ps)
jitter_b = np.random.normal(0, jitter_sigma_ps)
delta_t_ps = abs(jitter_a - jitter_b)

if not config.HOM_VISIBILITY_AT_MAX:
    visibility = calculate_visibility(delta_f_ghz, delta_t_ps)
else:
    visibility = 1
\end{lstlisting}
\end{figure*}

Whilst Listing 6 shows the code to convert $\Delta_{f}$ and $\Delta_{t}$ to $\sigma_{f}$ and $\sigma_{t}$ and then calculate the overall HOM visibility, $V_{HOM}$.

\begin{figure*}[h]
\begin{lstlisting}[caption={Calculate the HOM visibility, $V_{HOM}$, of the qubits recieved at the beam splitter.},label=HOMvisibility2, basicstyle=\fontsize{8}{10}\selectfont\ttfamily]
def calculate_visibility(delta_f_ghz: float, delta_t_ps: float) -> float:
    """Calculates the total HOM visibility from spectral and temporal mismatch."""
    vis_spectral = 1.0
    if config.PHOTON_FWHM_GHZ > 0:
        sigma_f = config.PHOTON_FWHM_GHZ / (2 * np.sqrt(2 * np.log(2)))
        vis_spectral = np.exp(-(delta_f_ghz**2) / (2 * sigma_f**2))

    vis_temporal = 1.0
    fwhm_duration_ps = (1000 / config.PHOTON_FWHM_GHZ) if config.PHOTON_FWHM_GHZ > 0 else float('inf')
    if fwhm_duration_ps > 0 and delta_t_ps > 0:
        sigma_t = fwhm_duration_ps / (2 * np.sqrt(2 * np.log(2)))
        vis_temporal = np.exp(-(delta_t_ps**2) / (2 * sigma_t**2))
    return vis_spectral * vis_temporal
\end{lstlisting}
\end{figure*}

\end{document}